\documentclass[useAMS,usenatbib,onecolumn]{mn2e}

\usepackage{dcolumn}
\usepackage{graphicx}
\usepackage{url}
\usepackage{color}
\usepackage{epstopdf}
\usepackage{textcomp}
\usepackage[T1]{fontenc}
\usepackage{longtable}

\newcommand{\ai}{{\it ab initio }}

\newcommand{\cm}{cm$^{-1}$}

\newcommand{\hato}{H$_2$$^{16}$O}

\newcommand{\octo}{H$_2$$^{18}$O}
\newcommand{\heto}{H$_2$$^{17}$O}

\newcommand{\s}{s$^{-1}$}
\newcommand{\gtot}{g$_{tot}$}

\def\a0{{$a_{\rm 0}$}}

\title[ExoMol XIX: H$_2$$^{18}$O and H$_2$$^{17}$O]{ExoMol molecular line lists XIX:
high accuracy computed hot line lists for H$_2$$^{18}$O and H$_2$$^{17}$O}
\date{\today}

\author[Polyansky et al]{
Oleg L. Polyansky$^{1,2}$, Aleksandra A. Kyuberis$^{2}$,  Lorenzo Lodi$^{1}$,
\newauthor Jonathan Tennyson$^{1}$\thanks{Email: j.tennyson@ucl.ac.uk}, Sergei N. Yurchenko$^{1}$,
 Roman I. Ovsyannikov$^{2}$, \newauthor  and Nikolai F. Zobov$^{2}$\\
$^1$Department of Physics and Astronomy, University College London, London WC1E 6BT, UK\\
$^{2}$Institute of Applied Physics, Russian Academy of Sciences,
Ulyanov Street 46, Nizhny Novgorod, Russia 603950.}
\date{Accepted XXXX. Received XXXX; in original form XXXX}

\pagerange{\pageref{firstpage}--\pageref{lastpage}} \pubyear{2014}
\begin{document}

\maketitle

\label{firstpage}

\begin{abstract}
  Hot line lists for two isotopologues of water, \octo\ and \heto, are
  presented.  The calculations employ newly constructed potential
  energy surfaces (PES) which take advantage of a novel method for
  using the large set of experimental energy levels for \hato\ to give high
  quality predictions for \octo\ and \heto.  This procedure greatly
  extends the energy range for which a PES can be accurately
  determined, allowing accurate prediction of higher-lying energy
  levels than are currently known from direct laboratory measurements.
  This PES is combined with a high-accuracy, {\it ab initio} dipole
  moment surface of water in the computation of all energy levels,
  transition frequencies and associated Einstein A coefficients for
  states with rotational excitation up to $J=50$ and energies up to
  30~000 \cm. The resulting HotWat78 line lists complement the
  well-used BT2 \hato\ line list (Barber et.al, 2006, MNRAS, {\bf
    368}, 1087).  Full line lists are made available in the electronic
  form as supplementary data to this article and at
  \url{www.exomol.com}.
\end{abstract}


\section{Introduction}

Water spectra can be observed from many different regimes in the
Universe, several of which are discussed further below. The
spectrum of water, particularly at elevated temperatures, is rich and
complex. A few years ago \citet{jt378} presented a comprehensive line
list, known as BT2, which used well-established theoretical procedures
to compute all the transitions of \hato\ of importance in objects with
temperatures up to 3000 K.  BT2 contains about 500 million lines. A
similar line list for HD$^{16}$O, known as VTT,
was subsequently computed by \citet{jt469}.

The BT2 line list has been extensively used. It forms the basis of
the most recent release of the HITEMP high-temperature spectroscopic database \citep{jt480}
and for the BT-Settl model \citep{BT-Settl} for stellar and substellar
atmospheres       covering the range from solar-mass stars to the
latest-type T and Y dwarfs.  BT2 has been used to detect and analyse
water spectra in objects as diverse as the Nova-like object V838 Mon
\citep{jt357}, atmospheres of brown dwarfs \citep{10RiBeMc.H2O} and
M subdwarfs \citep{14RaReAl.H2O}, and extensively for exoplanets
\citep{jt400,13BiDeBr.exo}. Within the solar system BT2 has been used
to show an imbalance between nuclear spin and rotational temperatures
in cometary comae \citep{jt330,jt349} and assign a new set of, as yet
unexplained, high energy water emissions in comets \citep{jt452}, as
well as to model water spectra in the deep atmosphere of Venus
\citep{Jeremy09}.

Although BT2 was developed for astrophysical use, it has been applied
to a variety of other problems including the calculation of the
refractive index of humid air in the infrared \citep{07Mathar.H2O},
high speed thermometry and tomographic imaging in gas engines and
burners \citep{07KrAnCa.H2O,10ReSa.H2O}, as the basis for an
improved theory of line-broadening \citep{jt431}, and to validate the
data used in models of the earths atmosphere and in particular
simulating the contribution of weak water transitions to the so-called
water continuum \citep{jt463}.

There are several water line lists published in
the literature \citep{jt197,97PaScxx.H2O,jt378,spectra}.
Two linelists have also been computed
 specifically for the isotopologues: \citet{jt438} created the 3mol 
room-temperature line lists for \hato,
\heto\ and \octo\ based on the PES                       of \citet{jt375};
Tashkun created a number of line lists based on the work of  \citet{97PaScxx.H2O}, see
\citet{spectra}. These are considered further below.

At present hot line lists are only published for \hato\ and
HD$^{16}$O.  However isotopically-substituted water containing
$^{18}$O or $^{17}$O provides important markers for a variety of
astronomical problems \citep{12NiGa.H2O}.  For example
\citet{14MaYaBa.H2O} recently detected \octo\ in the emission-line
spectrum of the luminous M-supergiant VY CMa.  Astronomical spectra of
water isotopologues \citep{13NeToAg.H2O} and their direct analysis in
cometary dust particles \citep{10FlStMe.H2O} and carbonaceous
chrondrites \citep{84ClMa.H2O,08VoHoBr.H2O} have been used to
determine formation mechanisms and constrain formation models.  Water
isotope ratios are also used to monitor stellar evolution
\citep{12AbPaBu.H2O} and to probe the atmosphere of Mars
\citep{15ViMuNo.H2O}. The seemingly minor isotopologues of water can
be important species in their own right with, for example, \octo\
being the fifth largest absorber of sunlight the earth's atmosphere.

There is therefore a need for line lists equivalent to BT2 for \heto\
and \octo\ to aid spectroscopic studies, and it is these that are
presented here. These lists form part of the ExoMol project \citep{jt528}
which aims to provide a comprehensive set of
molecular line lists for studies of
molecular line lists for exoplanet and other hot atmospheres.

Although our new line lists in some way mimic BT2, they
also take advantage of a number of recent theoretical developments. In
particular a IUPAC task group \citep{jt562} used a systematic
procedure \citep{jt412} to derive empirical energy levels for all the
main isotopologues of water \citep{jt454,jt482,jt539,jt576}. These
levels are combined with a newly-developed procedure for enhancing the accuracy of
calculations on isotopically substituted species, which
is used for the first time here. This ensures that most of the key
frequencies  in our new line lists are determined with an
accuracy close to experimental, even though many of them are yet to
be observed. Furthermore, theoretical work on improving the
accuracy and representation of the water dipole moment \citep{jt424,jt509}
has improved the accuracy with which water transition intensities
are predicted \citep{jt467}. Some of these advances have already
been used to create improved room temperature line lists for \heto\ and
\octo\ \citep{jt522} which were included in their entirety in the
 2012  release of HITRAN \citep{jt557s}.

The paper is structured as follows: section 2 outlines our overall
methodology and presents the derivation of potential energy surfaces
(PES).  The details of the
calculation of the new line lists,
along with comparison with
previous line lists, are given in section 3.  Section 4
discusses further improvement of the line list by  the substitution of
calculated energy levels with empirical ones,
together with the procedure used to
label energy levels with approximate vibrational and rotational
quantum numbers. Our results are discussed in section 5.

\section{Potential Energy Surfaces}

The fitting of water (\hato) PESs to experimental spectroscopic 
data has a long
history. The first fitted PES giving near to experimental accuracy was
PJT1 \citep{jt150}.  \citet{97PaScxx.H2O} constructed a fitted PES
starting from a highly accurate \ai\ calculation; all subsequent water
potentials followed this procedure and have been based on \ai\ studies
of increasing sophistication. As a result there are several very good
water PESs available \citep{jt308,jt438,11BuPoZo.H2O}.

Here we need a PES which  satisfies two criteria.  First, it should
be at least as accurate as the PES used for the BT2 line list with
the calculated energies ranging up to 30 000 \cm.  Second, the PES should be
adapted to the calculation of energy levels of the two water
isotopologues \heto\ and \octo. This second requirement is harder to
fulfill, as the characterisation of the experimental energy levels of
both \heto\ and \octo\ is significantly less extensive than for \hato\
\citep{jt562}.

To take advantage of the accumulated knowledge on the spectrum \hato\
in constructing a PES for \heto\ and \octo\ and following previous work
\citep{jt186,jt469,11BuPoZo.H2O}, we decided to fit a
Born-Oppenheimer (BO) mass-independent PES to the available data for
\hato\ and fix the adiabatic BO diagonal correction (BODC),
mass-dependent surface to the \ai\ value of \citet{jt309}. Obviously this
procedure requires the accuracy of predictions for \heto\ and \octo\
to be verified.  This is done by comparing the calculated \heto\ and \octo\
energy levels to the available experimentally-determined ones 
\citep{jt454,jt482}. 

We used the same fitting procedure as \citet{11BuPoZo.H2O}. Nuclear
motion calculations were performed with \textsc{DVR3D} \citep{jt338}.
As elsewhere, in the fit the experimentally derived energies of \hato\
for the $J=0,2$ and 5 rotational states by \citet{jt539} were used.

In the following our new empirical PES obtained using the fitting
procedure described above will be referenced to as PES1, while the PES
by \citet{11BuPoZo.H2O} will be referenced to as PES2.
Tables~\ref{PES17} and \ref{PES18} present a comparison between the $J
= 0$ energy levels calculated using PES1, PES2 for \heto\ and \octo\
respectively.  For comparison as a third column we present the $J = 0$
levels and corresponding discrepancies using the PES (called PES3 in
the tables) due to \citet{97PaScxx.H2O} taken from the linelist
calculated by Dr. S.A. Tashkun and summarised by \citet{spectra}.  The
line list based on PES3 was calculated for three temperatures: T=296
K, 1000 K and 3000 K. For all versions the highest value of the
rotational quantum number $J$ considered is 28 and the spectral range
is 0-28500 \cm. The number of lines for \octo\ is 108~784 and for
\heto\ 109~083.

  Indeed, one can see
that the agreement with the experiment is very good. 
Although the results obtained using PES2 are
somewhat better  than those for PES1. However employing PES1 gives us 
the opportunity to use the
information on \hato\ experimental energy levels to predict very
accurately energy levels of \heto\ and \octo. 
 We call these
predicted levels pseudo-experimental energies for the reasons explained below.
Table ~\ref{sd}
illustrates the unprecedented accuracy of the prediction of the \heto\ energy
levels for those states whose energies are known experimentally.
The slightly less good, but still very accurate, energy levels
predicted for \octo\ are shown in the column 2 of Table~\ref{sd}.
We might expect a similar level of accuracy for predictions of the
\heto\ and \octo\ energy levels for states yet to be measured for these
isotopologues, but known for \hato. We note that the standard
deviations given in Table~\ref{sd} are rather systematic suggesting
that further improvement in the predictions may be possible.
This and details of our final pseudo-experimental energy levels are discussed in
section 4.

\begin{table}
\caption{Comparison of calculated $J=0$ term values for \heto\
using three potentials  with experimental data.
Experimental (obs) data is taken from \citet{jt454}. }\label{PES17}
\begin{tabular}{crrrrrrrrrr}
\hline \hline
 $v_{1}$ & $v_{2}$ & $v_{3}$  & Observed &  PES1 & Obs.-Calc. & PES2 & Obs.-Calc. & PES3 & Obs.-Calc.\\
\hline

0	&	0	&	1	&	3748.318	&	3748.334	&	-0.02	&	3748.326	&	-0.01	&	3748.463	&	-0.15	\\
0	&	0	&	2	&   7431.076	&	7431.103	&	-0.03	&	7431.059	&	0.02	&	7431.467	&	-0.39	\\
0	&	0	&	3	&	11011.883	&	11011.936	&	-0.05	&	11011.860	&	0.02	&	11012.268	&	-0.38	\\
0	&	1	&	0	&	1591.326	&	1591.297	&	0.03	&	1591.342	&	-0.02	&	1591.413	&	-0.09	\\
0	&	1	&	1	&	5320.251	&	5320.241	&	0.01	&	5320.251	&	0.00	&	5320.378	&	-0.13	\\
0	&	1	&	2	&	8982.869	&	8982.868	&	0.00	&	8982.844	&	0.03	&	8983.118	&	-0.25	\\
0	&	1	&	3	&	12541.227	&	12541.267	&	-0.04	&	12541.207	&	0.02	&	12541.614	&	-0.39	\\
0	&	2	&	0	&	3144.980	&	3144.934	&	0.05	&	3144.993	&	-0.01	&	3145.085	&	-0.10	\\
0	&	2	&	1	&	6857.273	&	6857.260	&	0.01	&	6857.266	&	0.01	&	6857.476	&	-0.20	\\
0	&	7	&	1	&	13808.273	&	13808.224	&	0.05	&	13808.371	&	-0.10	&	13809.171	&	-0.90	\\
1	&	0	&	0	&	3653.142	&	3653.147	&	0.00	&	3653.121	&	0.02	&	3653.193	&	-0.05	\\
1	&	0	&	1	&	7238.714	&	7238.773	&	-0.06	&	7238.726	&	-0.01	&	7238.932	&	-0.22	\\
1	&	0	&	2	&	10853.505	&	10853.545	&	-0.04	&	10853.504	&	0.00	&		-       &	 -   	\\
1	&	0	&	3	&	14296.280	&	14296.340	&	-0.06	&	14296.265	&	0.01	&	14296.584	&	-0.30	\\
1	&	1	&	0	&	5227.706	&	5227.691	&	0.01	&	5227.704	&	0.00	&	5227.881	&	-0.18	\\
1	&	1	&	1	&	8792.544	&	8792.578	&	-0.03	&	8792.546	&	0.00	&	8792.816	&	-0.27	\\
1	&	2	&	0	&	6764.726	&	6764.747	&	-0.02	&	6764.722	&	0.00	&	6764.905	&	-0.18	\\
1	&	2	&	1	&	10311.202	&	10311.247	&	-0.05	&	10311.199	&	0.00	&	10311.421	&	-0.22	\\
1	&	3	&	1	&	11792.822	&	11792.861	&	-0.04	&	11792.834	&	-0.01	&	11793.172	&	-0.35	\\
2	&	0	&	0	&	7193.246	&	7193.265	&	-0.02	&	7193.257	&	-0.01	&	7193.394	&	-0.15	\\
2	&	0	&	1	&	10598.476	&	10598.550	&	-0.07	&	10598.483	&	-0.01	&	10598.763	&	-0.29	\\
2	&	1	&	1	&	12132.993	&	12133.056	&	-0.06	&	12132.984	&	0.01	&	12132.365	&	0.63	\\
2	&	2	&	1	&	13631.500	&	13631.542	&	-0.04	&	13631.489	&	0.01	&	13631.650	&	-0.15	\\
3	&	0	&	1	&	13812.158	&	13812.215	&	-0.06	&	13812.170	&	-0.01	&	13812.394	&	-0.24	\\
3	&	2	&	1	&	16797.168	&	16797.182	&	-0.01	&	16797.177	&	-0.01	&	16797.011	&	0.16	\\
4	&	0	&	1	&	16875.621	&	16875.662	&	-0.04	&	16875.643	&	-0.02	&	16875.474	&	0.15	\\

\hline
\end{tabular}
\end{table}

\begin{table}
\caption{Comparison of calculated $J=0$ term values for \octo\
using three potentials  with experimental data.
Experimental (obs) data is taken from \citet{jt454}. } \label{PES18}
\begin{tabular}{crrrrrrrrrr}
\hline \hline
 $v_{1}$ & $v_{2}$ & $v_{3}$  & Observed &    PES1  &  Obs.-Calc.  &  PES2  &  Obs.-Calc.  &  PES3  &  Obs.-Calc.  \\
\hline

0	&	0	&	1	&	 3741.57	&	 3741.581	&	-0.01	&	 3741.567	&	0.00	&	3741.575	&	-0.01	\\
0	&	0	&	2	&	 7418.72	&	 7418.741	&	-0.02	&	 7418.693	&	0.03	&	7418.759	&	-0.03	\\
0	&	0	&	3	&	10993.68	&	10993.734	&	-0.05	&	10993.659	&	0.02	&	10993.689	&	-0.01	\\
0	&	1	&	0	&	 1588.28	&	 1588.240	&	 0.04	&	 1588.271	&	0.00	&	1588.299	&	-0.02	\\
0	&	1	&	1	&	 5310.46	&	 5310.443	&	 0.02	&    5310.438	&	0.02	&	5310.388	&	0.07	\\
0	&	1	&	2	&	 8967.57	&	 8967.552	&	 0.01	&	 8967.519	&	0.05	&	8967.491	&	0.07	\\
0	&	1	&	3	&	12520.12	&	12520.153	&	-0.03	&	12520.089	&	0.03	&	12520.068	&	0.06	\\
0	&	2	&	0	&	 3139.05	&	 3138.999	&	 0.05	&	 3139.038	&	0.01	&	3139.031	&	0.02	\\
0	&	2	&	1	&	 6844.60	&	 6844.580	&	 0.02	&	 6844.566	&	0.03	&	6844.539	&	0.06	\\
0	&	2	&	2	&	10483.22	&	10483.264	&	-0.04	&	10483.202	&	0.02	&	10483.212	&	0.01	\\
0	&	3	&	0	&	 4648.48	&	 4648.435	&	 0.04	&	 4648.469	&	0.01	&	4648.452	&	0.03	\\
0	&	3	&	1	&	 8341.11	&	 8341.109	&	 0.00	&	 8341.086	&	0.02	&	8341.114	&	-0.01	\\
0	&	3	&	2	&	11963.54	&	11963.580	&	-0.04	&	11963.507	&	0.03	&	11963.615	&	-0.08	\\
0	&	4	&	0	&	 6110.42	&	 6110.408	&	 0.02	&	 6110.433	&	-0.01	&	6110.410	&	0.01	\\
0	&	4	&	1	&	 9795.33	&	 9795.354	&	-0.02	&	 9795.324	&	0.01	&	9795.329	&	0.00	\\
1	&	0	&	0	&	 3649.69	&	 3649.688	&	 0.00	&	 3649.649	&	0.04	&	3649.667	&	0.02	\\
1	&	0	&	1	&	 7228.88	&	 7228.934	&	-0.05	&	 7228.883	&	0.00	&	7228.888	&	0.00	\\
1	&	0	&	2	&	10839.96	&	10839.986	&	-0.03	&	10839.942	&	0.01	&		-       &	  - 	\\
1	&	0	&	3	&	14276.34	&	14276.389	&	-0.05	&	14276.318	&	0.02	&	14276.229	&	0.11	\\
1	&	1	&	0	&	 5221.24	& 	 5221.233	&	 0.01	&	 5221.227	&	0.02	&	5221.298	&	-0.05	\\
1	&	1	&	1	&	 8779.72	&	 8779.747	&	-0.03	&	 8779.707	&	0.01	&	8779.722	&	0.00	\\
1	&	1	&	2	&	12372.71	&	12372.723	&	-0.02	&	12372.679	&	0.03	&		-       &	  -  	\\
1	&	2	&	0	&	 6755.51	&	 6755.528	&	-0.02	&	 6755.483	&	0.03	&	6755.501	&	0.01	\\
1	&	2	&	1	&	10295.63	&	10295.673	&	-0.04	&	10295.616	&	0.02	&	10295.524	&	0.11	\\
1	&	3	&	0	&	 8249.04	&	 8249.063	&	-0.03	&	 8249.023	&	0.01	&	8249.073	&	-0.04	\\
1	&	3	&	1	&	11774.71	&	11774.742	&	-0.03	&	11774.701	&	0.01	&	11774.670	&	0.04	\\
2	&	0	&	0	&	 7185.88	&	 7185.894	&	-0.02	&	 7185.879	&	0.00	&	7185.880	&	0.00	\\
2	&	0	&	1	&	10585.29	&	10585.357	&	-0.07	&	10585.292	&	-0.01	&	10585.300	&	-0.01	\\
2	&	0	&	2	&	14187.98	&	14188.069	&	-0.09	&	14187.985	&	0.00	&		-       &	  -     \\
2	&	1	&	0	&	 8739.53	&	 8739.530	&	 0.00	&	 8739.520	&	0.01	&	8739.589	&	-0.06	\\
2	&	1	&	1	&	12116.80	&	12116.851	&	-0.05	&	12116.778	&	0.02	&	12116.833	&	-0.04	\\
2	&	2	&	0	&	10256.58	&	10256.604	&	-0.02	&	10256.569	&	0.02	&	10256.537	&	0.05	\\
2	&	2	&	1	&	13612.71	&	13612.745	&	-0.04	&	13612.688	&	0.02	&	13612.468	&	0.24	\\
2	&	3	&	0	&	11734.53	&	11734.543	&	-0.02	&	11734.517	&	0.01	&	11734.625	&	-0.10	\\
3	&	0	&	0	&	10573.92	&	10573.955	&	-0.04	&	10573.927	&	-0.01	&	10573.898	&	0.02	\\
3	&	0	&	1	&	13795.40	&	13795.455	&	-0.06	&	13795.410	&	-0.01	&	13795.280	&	0.12	\\
3	&	1	&	0	&	12106.98	&	12107.025	&	-0.05	&	12106.974	&	0.00	&	12107.006	&	-0.03	\\
3	&	2	&	1	&	16775.38	&	16775.396	&	-0.01	&	16775.385	&	0.00	&	16774.779	&	0.60	\\
4	&	0	&	1	&	16854.99	&	16855.126	&	-0.14	&	16855.099	&	-0.11	&	16854.534	&	0.46	\\

\hline
\end{tabular}
\end{table}

Recently, highly lying  energy levels of \octo\ have been measured using
multiphoton spectroscopy \citep{15MaKoZo}.  These levels lie at about
27 000 \cm\ and therefore provide a stringent test of our procedure.
                The highest upper energy level considered in this
work, as for BT2, is 30 000 \cm; Table~\ref{makarov} illustrates the
high quality of our calculations over the whole range considered.
In fact recent studies confirm that BT2 is not so accurate for these high energy states \citep{jt645}.

\begin{table}
\caption{Standard deviation in \cm\ with which our pseudo-experimental energy levels the of \heto\  and \octo\ predicted the observed ones
compiled by  \citet{jt482} as a function of rotational state, $J$, 
$N$ is number of levels used for calculation of the standard deviation.} \label{sd}
\begin{tabular}{ccccc}
\hline \hline
$J$ & $ N $ & {\heto} & $ N $ & {\octo} \\
\hline
0       &  27 & 0.0058  &  39 & 0.0092  \\
1       &  93 & 0.0056  & 124 & 0.0093  \\
2       & 161 & 0.0071  & 212 & 0.0109  \\
3       & 199 & 0.0074  & 254 & 0.0090  \\
4       & 236 & 0.0118  & 316 & 0.0147  \\
5       & 232 & 0.0103  & 335 & 0.0141  \\
6       & 263 & 0.0100  & 401 & 0.0116  \\
7       & 222 & 0.0138  & 385 & 0.0140  \\
8       & 182 & 0.0146  & 381 & 0.0130  \\
9       & 138 & 0.0123  & 335 & 0.0174  \\
10      & 116 & 0.0130  & 288 & 0.0176  \\
11      &  72 & 0.0080  & 232 & 0.0168  \\
12      &  47 & 0.0111  & 188 & 0.0201  \\
13      &  26 & 0.0083  & 135 & 0.0179  \\
14      &   9 & 0.0096  & 106 & 0.0198  \\
15      &   3 & 0.0150  &  73 & 0.0176  \\
16  &   1 & 0.0066  &  46 & 0.0184  \\
17      &   1 & 0.0015  &  19 & 0.0156  \\
18  &     &         &  11 & 0.0187 \\
\hline
\end{tabular}
\end{table}

\begin{table}
\caption{ Prediction of experimental energy levels  of \octo.
Experimental (obs) data is taken from \citet{15MaKoZo}.} \label{makarov}
\begin{tabular}{crrrrrr}
\hline \hline
$ J $ & Observed & Calculated & Obs.-Calc. \\

0	& 27476.33	& 27476.24	&  0.09 \\
1	& 27497.03	& 27496.92	&  0.11 \\
1   & 27510.64	& 27510.31	&  0.33 \\
1	& 27517.09	& 27517.44	& -0.35 \\
2	& 27537.12	& 27536.96	&  0.16 \\
2	& 27546.82	& 27546.45	&  0.37 \\
1	& 27509.55	& 27509.19	&  0.36 \\
2	& 27545.66	& 27545.28	&  0.38 \\

\hline
\end{tabular}
\end{table}

Thus, the line lists, details of whose  calculations are given in the
following section, are computed using a higher quality PES than that
used to compute BT2.  Three sets of energy levels are provided as part
of this line list. The first set is the variationally calculated energy
levels obtained using PES2. The second set comprises these energy levels
substituted by the experimental values \citep{jt454} where available. The third set is further with
pseudo-experimental energy levels substituted whenever \hato\ experimental energy
levels \citep{jt539} are available (see below). This third set is the one we recommend
for creating spectra with HotWat78 because of its increased accuracy.

\begin{figure}
\begin{center}
\includegraphics[angle=0, width=0.7\textwidth]{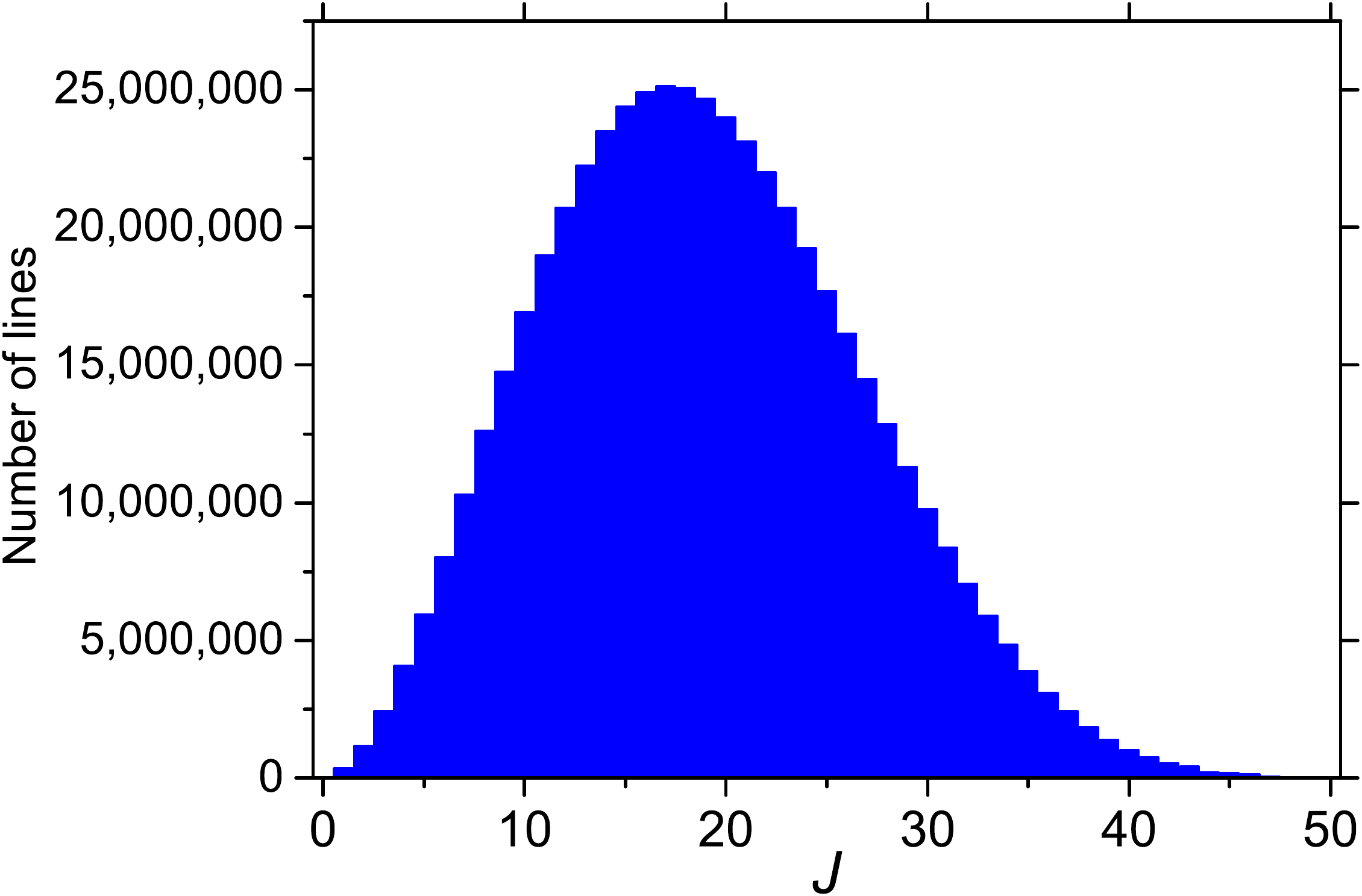}
\caption{The distribution of the \octo\ transitions per $J$ in the line HotWat78 list. }
\label{f:18:lines:J}
\end{center}
\end{figure}

\begin{figure}
\begin{center}
\includegraphics[angle=0, width=1.065\textwidth,
height=0.18\textheight]{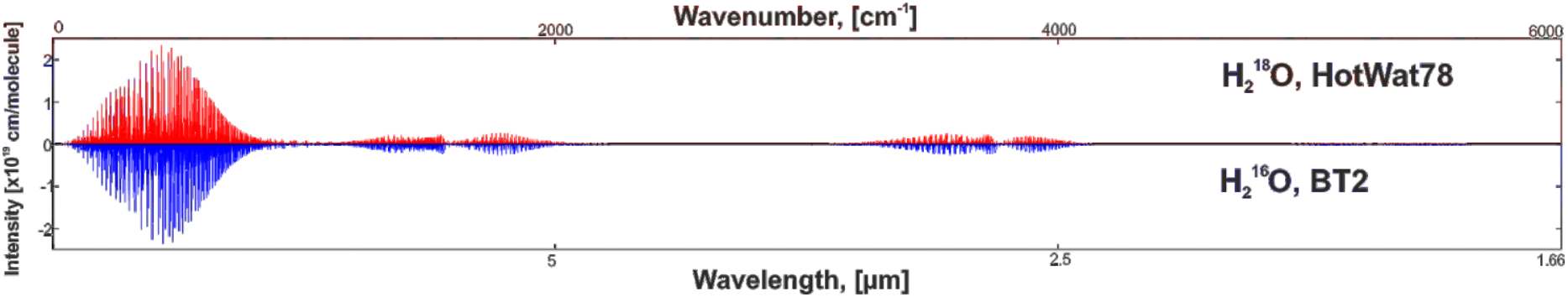}
\includegraphics[angle=0,  width=1.07\textwidth]{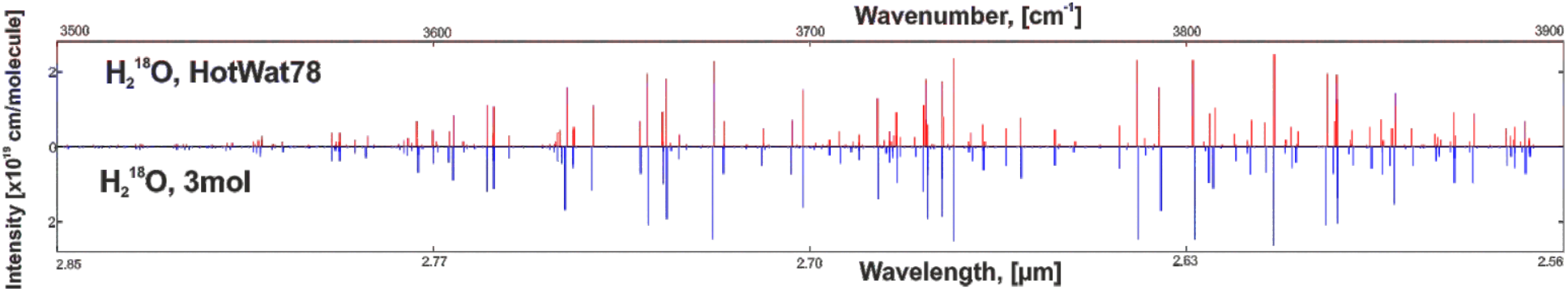}
\includegraphics[angle=0, width=1.15\textwidth,
height=0.18\textheight]{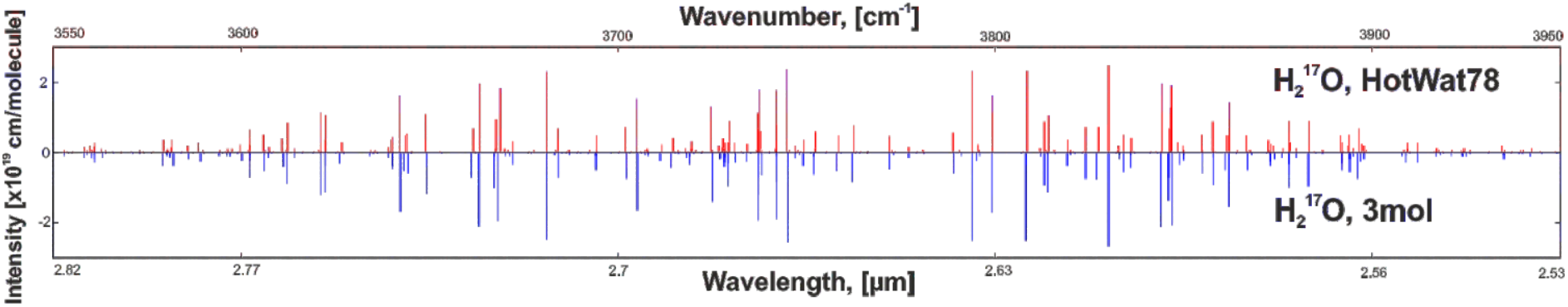}
\includegraphics[angle=0, width=1.09\textwidth,
height=0.18\textheight]{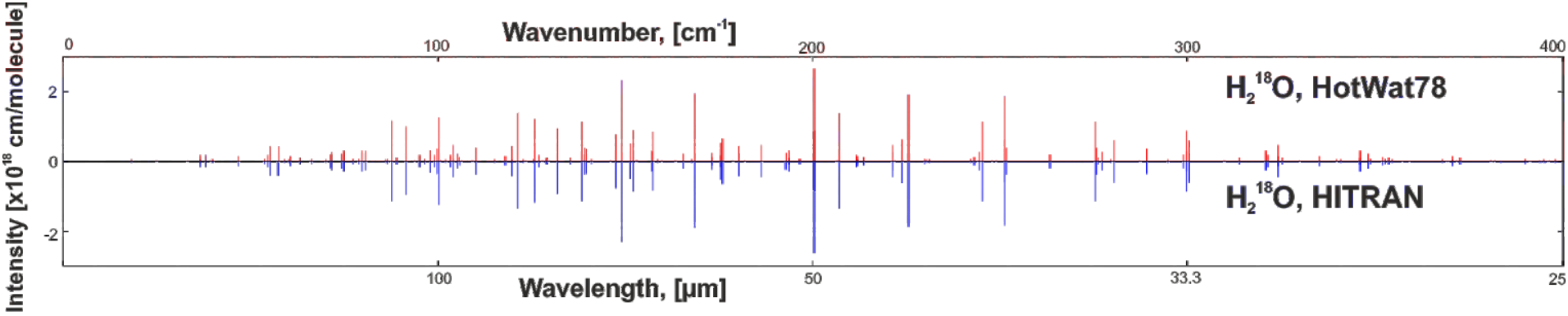}
\includegraphics[angle=0, width=1.14\textwidth]{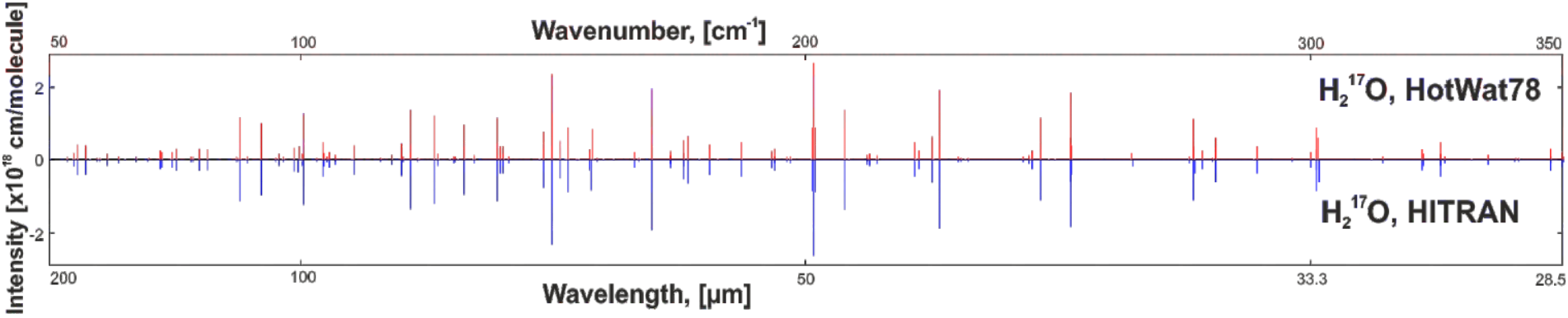}
\caption{Comparison  between BT2 and HotWat78 for \octo\ at the temperature
$T$=2000~K, and 
comparison of HotWat78 with 3mol \citep{jt438} and HITRAN at  $T$=296~K 
for \octo\ and \heto\ respectively.}
\label{Figure01}
\end{center}
\end{figure}

\begin{figure}
\begin{center}
\includegraphics[angle=0, width=1.06\textwidth,
height=0.2\textheight]{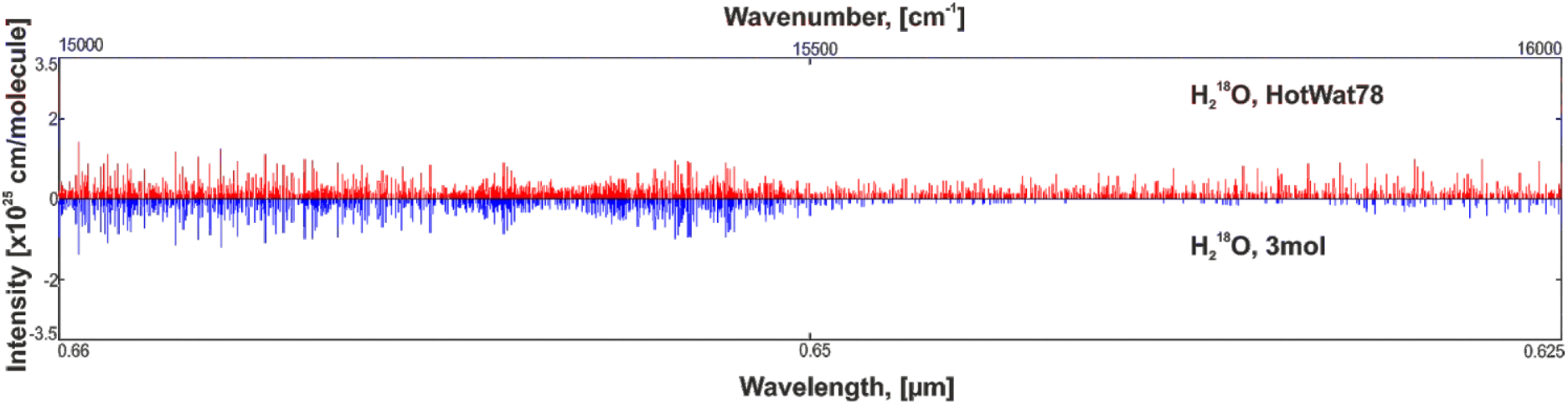}
\includegraphics[angle=0, width=1.06\textwidth,
height=0.2\textheight]{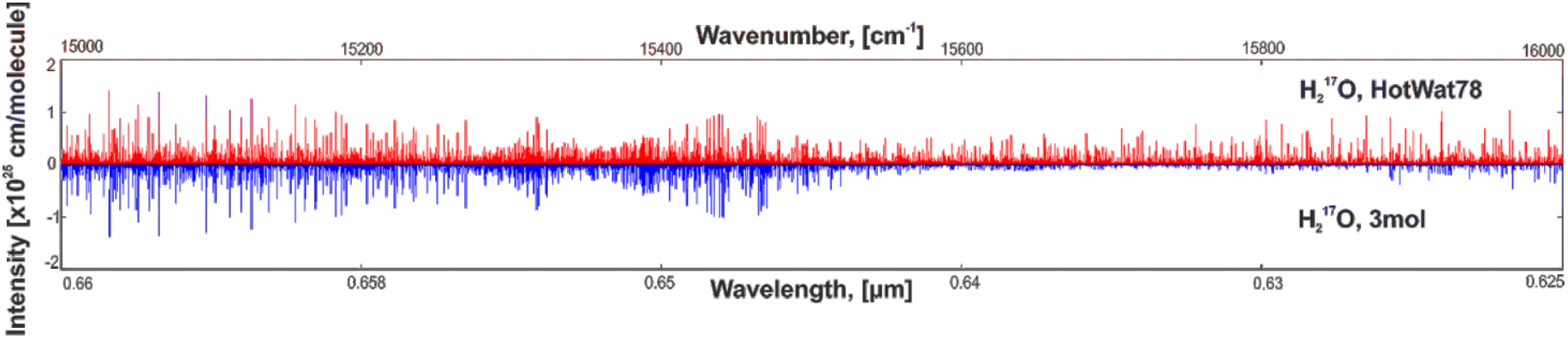}
\caption{Comparison of \octo\ and \heto\ between 3mol \citep{jt438} and
HotWat78 at the temperature $T$=3000~K.}
\label{Figure02}
\end{center}
\end{figure}

\section{Line list calculations for \heto\ and \octo}

The line list calculations were performed with the \textsc{DVR3D} program
        suite \citep{jt338} using the PES1 and PES2 discussed above, and the {\it ab initio}
dipole moment surfaces LTP2011S of \citet{jt509}. As for BT2, the highest
rotational state, $J$, in the calculation was taken as $J=50$ and the
limiting energy as 30 000 \cm. Analysis using the \hato\ partition
function \citep{jt263} performed in BT2 suggests that these parameters are
sufficient to cover all transitions longwards of 0.5 $\mu$m for
temperatures up to 3000 K.

Wavefunctions were obtained by solving the nuclear Schr\"{o}dinger
equation using two-step procedure of calculation of rovibrational
energies \citep{jt46}. The calculations benefitted from recent algorithmic
improvements \citep{jt626}, in particular
 in the method used to construct the final Hamiltonian
matrices for $J>0$ due to \citet{jt640}.  Transition intensities
were computed for $\Delta J = 0$ and 1 for all four symmetries and every
$J\le 50$.  The matrix elements of the DMS were calculated using the program
\textsc{Dipole} of the suite \textsc{DVR3D} and the actual spectrum for both
isotopologues was generated with the program Spectra. About 500 million
transitions were calculated for each isotopologue.

Figure~\ref{f:18:lines:J} shows the distribution of the \octo\ lines in HotWat78.

Using our calculations we provide the values of partition function for both
isotopologues for wide range of temperatures,
which are presented in the Table~\ref{pf} as well as in the supplementary data
on a grid of 1~K. 
We use the HITRAN convention \citep{03FiGaGo.partfunc} and include the nuclear
statistical weights $g_{\rm ns}$ 
in to the partition function explicitly \citep{jt631}. The nuclear statistical
weights for \octo\ are the same 
as for the main isotopologue, 1 and 3 for the para- and ortho- states, 
respectively. In case of \heto, $g_{\rm ns}$  are 6 (para) and 18 (ortho). 
For calculation of partition functions for \octo\ and \heto\ we used all available energy levels 
with applying the cut-off at 30000 \cm. 

\begin{table}
\caption{Partition Function of \heto\ and \octo.}\label{pf}
\begin{tabular}{rrrr}
\hline \hline
$T(K)$ & {\heto} & {\octo}\\

\hline

10	    &	    7.97970859	&	   1.33135007	  \\    
20	    &	    20.1629004	&	   3.37074465 	  \\
40	    &	    56.7292812	&	   9.48860674	  \\
60	    &	    101.331587	&	   16.9509639 	  \\
80	    &	    153.237432	&	   25.6357152	  \\
100	    &	    211.822453	&	   35.4382143	  \\
200	    &	   	587.053283  &	   98.2237727     \\
296	    &	  	1052.12202  &	   176.043783     \\
300	    & 	  	1073.45356  &	   179.613285	  \\
400	    &	  	1654.78625  &	   276.895547	  \\
500	    &	    2328.51505	&	   389.655412	  \\
600 	&	    3099.26294	&	   518.674912	  \\
800 	&	    4966.65892	&	   831.352302	  \\
1000	&	    7346.85187	&	   1230.02825	  \\
1200	&	    10357.5304	&	   1734.46724     \\
1400	&	 	14140.2160  &	   2368.43292	  \\
1500	&	    16371.1820	&	   2742.40404	  \\
1600	&	   	18857.9004  &	   3159.29345     \\
1800	&	    24694.5428	&	   4137.93895	  \\
2000	&	 	31855.8230  &	   5338.90908	  \\
2200	&	 	40570.4778  &	   6800.61746	  \\
2400	&	 	51091.7815  &	   8565.59949	  \\
2500	&	 	57116.1119  &	   9576.29200	  \\
2600	&	 	63698.8388  &	   10680.7274     \\
2800	&	 	78697.3411  &	   13197.3344	  \\
3000	&	 	96419.4218  &	   16171.1873	  \\
3200	&		117222.299  &	   19662.2543	  \\
3400	&		141485.523  &	   23734.2409	  \\
3500	&		155038.487  &	   26008.8411	  \\
3600	&		169606.832  &	   28453.8904	  \\
3800	&		201996.792  &	   33890.0829	  \\
4000	&		239072.534  &	   40112.7834	  \\
4200	&		281250.969  &	   47191.9028	  \\
4400	&		328941.890  &	   55196.1417     \\
4500	&		354979.000  &	   59566.0429     \\
4600	&		382541.321  &	   64191.8753     \\
4800	&	    442425.403	&	   74242.1299     \\
5000	&	    508945.054	&	   85405.6885	  \\
5200	&		582421.516  &	   97736.3470     \\
5400	&		663142.877  &	   111282.333     \\
5500	&		706300.716  &	   118524.515     \\
5600	&		751361.549  &	   126085.883	  \\
5800	&		847292.676  &	   148990.861 	  \\
6000	&		951113.377  &	   159603.233	  \\

\hline
\end{tabular}
\end{table}

\section{Improved  pseudo-experimental energy levels}
The series of IUPAC papers on the various isotopologues of water
\citep{jt454,jt482,jt539,jt576} used measured transition frequencies
to derive ro-vibrational energy levels using the so-called MARVEL 
(measured active rotation-vibration energy levels)
procedure \citep{jt412,12FuCs.method}. These energy levels can be used to generate
pseudo-experimental values of the line frequencies in our line lists
when the calculated energy level is substituted by the corresponding
(pseudo-)experimental one. The comparison of these generated line
frequencies with actual experimental ones demonstrate near-perfect
coincidence. The number of generated pseudo-experimental lines is
significantly higher than the number of the directly observed
lines because line frequencies between pseudo-experimental levels can
be predicted to high accuracy even when the lines have not been measured,
as demonstrated by \citet{jt539}. Less than 200 000 experimentally
observed \hato\ lines give rise to about 5 000 000 lines with
pseudo-experimental
frequencies generated in this way. Use of such a procedure provides
significantly more accurate line lists than just the calculated ones.
We therefore substituted our computed energy with those of
\citet{jt454} where possible.

However as described in section 2, the procedure for fitting
PES using \hato\ data opens the way for us to further improve the accuracy of
the calculated line lists.  Looking at Table \ref{oc}, we
can see that the
obs$-$calc residuals for a particular \hato\ vibrational state are very
similar to the residuals for the same states of \heto\ and \octo. The
following procedure can be used to exploit this.  First let us
consider the idealised situation when all the residuals for energy
levels of \hato, $R_{v,J}(16)$, are exactly equal to those of \octo,
$R_{v,J}(18)$, where $(v,J)$ represent the vibrational and rotational
quantum numbers. In this case we can predict the precise ``estimated''
value of an \octo\ level, $E_{v,J}^{\rm est}(18)$, from the
empirically-determined levels of \hato, $E_{v,J}^{\rm obs}(18)$
\begin{equation}
E_{v,j}^{\rm est}(18) = E_{v,J}^{\rm calc}(18) + R_{v,J}(18) = E_{v,J}^{\rm calc}(18) + R_{v,J}(16)
\end{equation}
where $E_{v,J}^{\rm calc}(18)$ is the corresponding calculated \octo\
energy level.  So even if the level of the \octo\ isotopologue has yet
to be observed, its pseudo-experimental value can be retrieved from
the calculated level of \octo\ using our calculations plus the
residual for \hato\ provided the experimental level of \hato\ is
known.

\begin{table}
\caption{      Vibrational band origins, in \cm, for \hato, \heto\  and \octo. Observed (obs)
data is taken from \citet{jt539} and \citet{jt454};  calculated results are
given as observed minus calculated (o--c).}  \label{oc}
\begin{tabular}{crrrrrr}
\hline \hline
$(v_1v_2v_3)$ & \hato& & \heto& & \octo\\
              & obs & o--c&  obs & o--c& obs & o--c\\
\hline
(010)  &  1594.75 &  0.019& 1591.33 &  0.028 & 1588.28 &  0.036 \\
(020)  &  3151.63 &  0.040& 3144.98 &  0.046  &3139.05&  0.051\\
(100)  &  3657.05 & -0.007& 3653.14 & -0.005  &3649.69 & -0.002\\
(110)  &  5234.97 &  0.005& 5227.71 & 0.014  &5221.24  &  0.010 \\
(120)  &  6775.09 & -0.028& 6764.73 & -0.022  &6755.51 & -0.018 \\
(200)  &  7201.54 & -0.024&7193.25 & -0.019  & 7185.88 & -0.016\\
(012)  &  9000.14 & -0.009&8982.87 & 0.001  &8967.57 &  0.013  \\
(102)  & 10868.88 & -0.049&10853.51 & -0.040  & 10839.96 & -0.030 \\
       &          &        &       &        &    &     \\
(001)  &  3755.93 & -0.017&3748.32 & -0.015  & 3741.57 & -0.014   \\
(011)  &  5331.27 & -0.002&5320.25 & 0.010  & 5310.46 &  0.019    \\
(021)  &  6871.52 &  0.004&6857.27 & 0.012  & 6844.60 &  0.019    \\
(101)  &  7249.82 & -0.063& 7238.71 & -0.059  & 7228.88 & -0.051    \\
(111)  &  8807.00 & -0.044&8792.54 & -0.034  & 8779.72 & -0.027    \\
(121)  & 10328.73 & -0.055&10311.20 & -0.045  & 10295.63 & -0.039    \\
(201)  & 10613.36 & -0.074&10598.48 & -0.075  & 10585.29 & -0.072    \\
(003)  & 11032.40 & -0.061&11011.88 & -0.053  & 10993.68 & -0.053   \\
(131)  & 11813.20 & -0.041&11792.82 & -0.039  & 11774.71 & -0.034    \\
(211)  & 12151.25 & -0.072&12132.99 & -0.064  & 12116.80 & -0.054    \\
(113)  & 12565.01 & -0.050&12541.23 & -0.041  & 12520.12 & -0.030    \\
(221)  & 13652.66 & -0.045&13631.50 & -0.042  & 13612.71 & -0.035    \\
(301)  & 13830.94 & -0.062&13812.16 & -0.057  & 13795.40 & -0.057   \\
(103)  & 14318.81 & -0.069&14296.28 & -0.061  & 14276.34 & -0.053    \\

\hline
\end{tabular}
\end{table}

         Table~\ref{oc} shows  that residuals for \hato\ and \octo\
are slightly different, we can therefore improve this procedure. We notice from
the Table~\ref{oc}, that the \hato\ and \octo\ residuals differ by similar
amounts. If we average this value:
\begin{equation}
 \Delta R(18) = \frac{1}{N}\sum_{v=1}^N R_{v,0}(18) - R_{v,0}(16).
\end{equation}
where $N$ runs over the number of vibrational states for which $J=0$ levels are known, which corresponds to
40 for \heto\ and 24 for \octo. Then we can use this
average difference to further correct our
estimated \octo\ energy levels using the revised formula:
\begin{equation}
E_{v,j}^{\rm est}(18)  = E_{v,J}^{\rm calc}(18) + R_{v,J}(16) + \Delta R(18).
\end{equation}
Calculating the observed values of energies of \octo\ using Eq.~(1)
gives a standard deviation for $E_{v,j}^{\rm est}(18)$ levels from the
known experimental values, $E_{v,j}^{\rm obs}(18)$, of 0.009 \cm.
However, $\Delta R(18)$ is 0.006 \cm. If instead we use Eq.~(3), then
the standard deviation reduces to 0.003 \cm.  Although $\Delta R(18)$
is evaluated for $J=0$ only,  this procedure still works for higher
$J$ values. For example it also results in a standard deviation of 0.003
\cm\ when applied to the $J=10$ levels of the (010) state.

This procedure, which can clearly also be applied to \heto, leads to
the generation of about 5 million transitions which involve the
pseudo-experimental levels of \heto\ and \octo. It therefore provides
a line list with much more accurate values of the
frequencies of these transitions: in general better by about 0.005 \cm\
for \heto\ and somewhat worse  for \octo,
 but still much more accurate than
possible with variational calculations.

The reason this procedure can be applied to the construction of the pseudo-experimental
values of the energy levels of minor isotopologues is that for the major water isotopologue
\hato\ the number of energy levels known experimentally is significantly higher, then that
for \heto\ and \octo. For example the assignment of weak \hato\ lines in various
regions is available \citep{jt360,jt218,jt285}, where isotopologues data are not known.
As a result very highly-excited bending \citep{jt203,jt362} and stretching energy
levels \citep{07MaMuZoSh,jt467,jt472} are known, which form the basis 
upon which our  pseudo-experimental
energy levels are constructed.

\section{Results}

The newly constructed \heto\ and \octo\ line lists are named HotWat78.
  The new HotWat78 line lists are
calculated for J $\leq$ 50 and for the spectral range 0-30000 \cm.
HotWat78 contains 519~461~789 lines for \octo\ is 519 461 789 and
513~112~779 lines for \heto.  The new linelist is both the most
complete and the most accurate one, see Tables ~\ref{PES17} and
~\ref{PES18}.
They are stored in the ExoMol format \citep{jt548} which uses the
compact storage of results originally developed for BT2.
This involves using a states file (\texttt{.states}), see Table~\ref{states}, and a transitions file (\texttt{.trans}),
see Table~\ref{trans}.
 The energy levels in the states files are marked as `observed' if the results are taken
from the IUPAC compilation, `estimated' if they are generated using
Eq.~(3) or as `calculated', for which the results of the PES2 calculation
are used.

\begin{table}
\caption{Extract from the final states file for \heto.}\label{states}
\begin{tabular}{crcccccccc}
\hline
$i$ &  $\tilde{E}$  & \gtot  & $J$ & Ka & Kc & v1 & v2 & v3 & $S$ \\
\hline
1  &     0.000000 & 6 & 0 & 0 & 0 & 0 & 0 & 0 & A1 \\
2  &  1591.322876 & 6 & 0 & 0 & 0 & 0 & 1 & 0 & A1 \\
3  &  3144.980225 & 6 & 0 & 0 & 0 & 0 & 2 & 0 & A1 \\
4  &  3653.145752 & 6 & 0 & 0 & 0 & 1 & 0 & 0 & A1 \\
5  &  4657.115211 & 6 & 0 & 0 & 0 & 0 & 3 & 0 & A1 \\
6  &  5227.703125 & 6 & 0 & 0 & 0 & 1 & 1 & 0 & A1 \\
7  &  6121.557129 & 6 & 0 & 0 & 0 & 0 & 4 & 0 & A1 \\
8  &  6764.726562 & 6 & 0 & 0 & 0 & 1 & 2 & 0 & A1 \\
9  &  7193.246582 & 6 & 0 & 0 & 0 & 2 & 0 & 0 & A1 \\
10 &  7431.093262 & 6 & 0 & 0 & 0 & 0 & 0 & 2 & A1 \\
11 &  7527.489258 & 6 & 0 & 0 & 0 & 0 & 5 & 0 & A1 \\
12 &  8260.781250 & 6 & 0 & 0 & 0 & 1 & 3 & 0 & A1 \\
13 &  8749.905273 & 6 & 0 & 0 & 0 & 2 & 1 & 0 & A1 \\
14 &  8853.288086 & 6 & 0 & 0 & 0 & 0 & 6 & 0 & A1 \\
15 &  8982.860352 & 6 & 0 & 0 & 0 & 0 & 1 & 2 & A1 \\
16 &  9708.538086 & 6 & 0 & 0 & 0 & 1 & 4 & 0 & A1 \\
17 & 10068.091797 & 6 & 0 & 0 & 0 & 0 & 7 & 0 & A1 \\
18 & 10269.661133 & 6 & 0 & 0 & 0 & 2 & 2 & 0 & A1 \\
19 & 10501.353516 & 6 & 0 & 0 & 0 & 0 & 2 & 2 & A1 \\
20 & 10586.049805 & 6 & 0 & 0 & 0 & 3 & 0 & 0 & A1 \\
\hline
\end{tabular}
\mbox{}\\
{\flushleft
$i$:   State counting number.     \\
$\tilde{E}$: State energy in \cm. \\
$g$: Total state degeneracy.\\
$J$: Total angular momentum            \\
$K_a$: Asymmetric top quantum number.\\
$K_c$:  Asymmetric top quantum number.\\
$\nu_1$:   Symmetric stretch quantum number. \\
$\nu_2$:   Bending quantum number. \\
$\nu_3$:   Asymmetric stretch quantum number. \\
$S$: State symmetry in C$_{2v}$.}

\end{table}

\begin{table}
\caption{Extract from the transitions file for \heto}\label{trans}
\begin{tabular}{ccccr}
\hline
$f$ & $i$ & $A_{fi}$  \\
\hline
142344  &    150189 & 5.6651e-05 \\
  2235  &      2362 & 1.7434e-03 \\
 34497  &     35342 & 5.7700e-09 \\
125681  &    114596 & 5.5394e-10 \\
135143  &    128340 & 6.3329e-08 \\
 24055  &     16736 & 1.5208e-03 \\
147918  &    137719 & 1.3405e-04 \\
 45027  &     45537 & 8.0306e-07 \\
 37457  &     31884 & 9.0168e-08 \\
 39192  &     43632 & 7.3676e-07 \\
 25153  &     26085 & 4.3393e-05 \\
131146  &    124272 & 8.5679e-04 \\
134840  &    128287 & 8.5680e-04 \\
 88744  &     94220 & 1.2221e-03 \\
102017  &    106580 & 2.4131e-04 \\
193489  &    187074 & 2.7697e-06 \\
202910  &    204558 & 7.0571e-03 \\
 53725  &     50906 & 1.8345e-06 \\
142862  &    135857 & 2.5908e-05 \\
\hline
\end{tabular}
\mbox{}\\
{$f$} : Upper state counting number.  \\
{$i$}: Lower state counting number. \\
$A_{fi}$ : Einstein-A coefficient in \s.\\

\end{table}

The states file lists all the ro-vibrational levels for each $J$ and for
four C$_{2v}$ symmetries. It is common to further label the every level with
(approximate) vibrational quantum numbers $(v_1, v_2, v_3)$ which
correspond to the symmetric stretch, bending and asymmetric stretch
modes, respectively and the Rotational levels within each vibrational state
by $J,K_a,K_c$, where again the projection quantum
numbers $K_a$ and $K_c$ are approximate.  \textsc{DVR3D} does not provide these
approximate  labels but there are several methods available for labeling water
energy levels
\citep{97PaScxx.H2O,12SzFaCs.method,jt438}. Here we label levels with $J\leq20$
and energies below 20~000 \cm.  As our energy levels differ by less
than 1 \cm\ from those of \citet{jt438}, transferring the labels from
this previous study proved to be straightforward. We note that the
labels we use are based on the normal modes from a harmonic oscillator model. It
is well know that the higher stretching states of water are better
represented with a local-mode model \citep{84ChHoxx}. However, there is a
one-to-one correspodance between the two labelling schemes
\citep{jt242}; the use of normal mode labels are used for simplicity.

The accuracy of the present line lists can be established by the
comparison with the previous line lists calculations. Two types of
comparison could be made.  The overall picture for the high
temperature is that the coverage the HotWat78 \heto\ and \octo\ line
lists should be very similar to BT2, but that both the predicted
intensives and the line positions should be significantly better.
Furthermore lines may shift by between a few \cm\ to a few tens of
\cm\ between isotopologues. Figure 1 demonstrate that, as
expected, the overall picture is very similar for
BT2 (\hato) and HotWat78 (\heto\ and \octo). 
Here we provide the comparison only for \octo\, but for the \heto\ it looks the same.

Figures 2 and 3 illustrate the similarity of the HotWat78 line lists with the previous high 
accuracy \heto\ and \octo\ line lists (called 3mol) of \citet{jt438} for these
molecules at the room temperature. 
Figures 4 and 5 also  provide a comparison
with the HITRAN data  for the room temperature for \heto\ and \octo.
These figures only provide an overview, but a detailed line by line
comparison confirms that all the calculations we present here are
done correctly.

The present line lists are significantly more complete, but this is
only apparant 
at higher temperatures, see Fig. 3.  For the room
temperature the previous line lists should look similar, as they
indeed do, see Figures 2.

\section{Conclusions}

This paper reports  hot line lists for  \heto\ and \octo.
These line lists represent significant improvement on both coverage and
accuracy of the previous \heto\ and \octo\ line lists \citep{spectra,jt438}.
 The predicted
frequencies in these line lists have been significantly improved
using information obtained from the corresponding \hato\ empirical
energy levels. This procedure can be adapted to give improved
predictions of energy levels and transition frequencies for isotopologues
of molecules for whom the empirical energy levels of the parent molecule
are well-known.

The complete HotWat78 line lists for \heto\ and \octo\ can be
downloaded from the CDS, via
\url{ftp://cdsarc.u-strasbg.fr/pub/cats/J/MNRAS/}, or via
\url{http://cdsarc.u-strasbg.fr/viz-bin/qcat?J/MNRAS/}.
The line lists
together with auxiliary data including the potential parameters,
dipole moment functions, and theoretical energy levels can be also
obtained at \url{www.exomol.com}, where they form part of the enhanced
Exomol database \citep{jt631}. The BT2 \hato\ line list
\citep{jt378} is already available from these sources.

Finally we note that pressure-broadening has been shown to have a significant
effect on water spectra in exoplanets \citep{jt521}. ExoMol, in common
with other databases, assumes that pressure-broadening parameters for
\heto\ and \octo\ are the same as those for \hato. This assumption is built
into the recently updated structure of the ExoMol database \citep{jt631}.
\citet{jt669} have recently presented a comprehensive set of pressure-broadening
parameters for \hato\ lines which form the basis for the ExoMol pressure-broadening
diet for water \citep{jtdiet}. These parameters, which are available
on the ExoMol website, are also suitable for use with the HotWat78 line lists.

\section*{Acknowledgements}

This work is supported by ERC Advanced Investigator Project 267219
and by the Russian Fund for Fundamental Studies.

\bibliographystyle{mn2e}

\end{document}